\documentclass[english,twocolumn]{aastex63}
\usepackage[T1]{fontenc}
\usepackage[latin1]{inputenc}
\usepackage{subfigure}
\usepackage{amssymb}
\usepackage{float}
\usepackage{graphicx}
\usepackage{amssymb}

\makeatletter

\usepackage{natbib}
\usepackage{bm}
\usepackage{amssymb}
\usepackage{amsmath,amsfonts,amsthm,bm} 
\usepackage{babel}
\usepackage{graphicx}

\newcommand{\eq}[1] { Eq.~(\ref{#1})}
\newcommand{\be}{\begin{equation}} 
\newcommand{\ee}{\end{equation}} 
\newcommand{\bea}{\begin{eqnarray}} 
\newcommand{\eea}{\end{eqnarray}}

\def\btau{{\bm{\tau}}}              %
\def\bu{{\bf u}}              %
\def\bR{{\bf R}}              %
\def\osigma{\overline{ \sigma}}              %
\def\br{{\bf r}}              %
\def\bg{{\bf g}}              %
\def\orho{\overline{\rho}}              %
\def\l{{\langle }}              %

\makeatother

\received{September 14, 2019}
\accepted{October 11, 2019}
\submitjournal{ApJL}

\begin{document}

\title{The interstellar object  'Oumuamua  as a fractal dust aggregate}

\correspondingauthor{Eirik G. Flekk{\o}y}
\email{e.g.flekkoy@fys.uio.no}

\author[0000-0002-0786-7307]{Eirik G. Flekk{\o}y}
\affiliation{PoreLab, SFF, the Njord Centre \\
Department of Physics, University of Oslo, P. O. Box 1048 \\
Blindern, N-0316, Oslo, Norway}

\author{Jane Luu }
\affiliation{Centre for Earth Evolution and Dynamics, Department of Geosciences, University of Oslo} 
\affiliation{Institute of Theoretical Astrophysics, University of Oslo}

\author{Renaud Toussaint}
\affiliation{Institut de Physique du Globe de Strasbourg, University of Strasbourg}
\affiliation{PoreLab, 2SFF, the Njord Centre \\
Department of Physics, University of Oslo, P. O. Box 1048 \\
Blindern, N-0316, Oslo, Norway}


\begin{abstract}
The first known interstellar object 'Oumuamua exhibited a nongravitational acceleration that appeared inconsistent with cometary outgassing, leaving  radiation pressure as the most likely force.  Bar the alien lightsail hypothesis, an ultra-low density due to a fractal structure might also explain the acceleration of 'Oumuamua by radiation pressure (Moro-Martin 2019).  In this paper we report a decrease in 'Oumuamua's rotation period based on ground-based observations, and show that this spin-down can be explained by the YORP effect if 'Oumuamua is indeed a fractal body with the ultra-low density of $10^{-2}$ kg m$^{-3}$.  We also investigate the mechanical consequences of 'Oumuamua as a fractal body subjected to rotational and tidal forces, and show that a fractal structure can survive these mechanical forces.
\end{abstract}

\keywords{editorials}

\section{ Introduction}

The first known interstellar object 1I/'Oumuamua was discovered by the Pan-STARRS telescope in October 2017, and after a short and intense observational campaign, it left behind far more questions than answers.  Ground-based observations (\cite{bannister2017}, \cite{jewitt2017}, \cite{meech2017}, \cite{bolin2017}, \cite{drahus2018}, \cite{fraser2018}, \cite{knight2017}) generally agreed on the following: 1) effective radius: 50 - 130 m, 2)  axis ratio: 6:1, and 3) a rotation period 6.8 - 8.7hr.  
However, the fact that Oumuamua could not be observed in the infrared spectrum (see  \cite{trilling18})
introduces a significant uncertainty in its  albedo, and hence its size.

There was much speculation about its origin, ranging from an isotropic background population (\cite{do2018})  to non-isotropic distributions  (\cite{moromartin18}, \cite{moromartin19a}) originating from a nearby young solar system  (\cite{gaidos17}).  Efforts to retrace 'Oumuamua's to its origin were complicated by the observed non-gravitational acceleration (\cite{micheli18}).  Micheli et al., as well as \cite{seligman19},  attributed the acceleration to cometary outgassing, in which case it could be explained by a dust mass loss rate of a few $\times$ 0.1 kg/s.  This  mass loss rate contradicts upper limits on the mass loss rate imposed by both ground- and space-based observations.  For example, \cite{jewitt2017} saw no coma in deep images of the object, and calculated an upper limit of $10^{-4}$ kg/s.   \cite{rafikov18} pointed out that outgassing torques should have spun up the object to the point of fragmenting due to rotation fission, but Oumuamua's rotation period appeared stable over the $\sim$ month-long observing campaign.  Finally, 'Oumuamua was not detected by the Spitzer Space Telescope(\cite{trilling18}), and the non-detection placed an upper limit on the CO and CO$_2$ production rate that was 2-3 orders of magnitude lower than that required for acceleration.  

\cite{bialy18} suggested instead that radiation pressure could explain the extra acceleration, only if the mass-to-area ratio were as small as $\sim 1$ kg m$^{-2}$, akin to that of a lightsail.   Finally, \cite{moromartin19b} pointed out that propulsion by radiation pressure could also be effective if 'Oumuamua had a highly porous fractal structure, with a density in the range $\sim 10^{-2}- 10^{-3}$ kg m$^{-3}$, depending on the fractal parameters.  The fractal hypothesis is  a natural one, as it has long been expected that random growth processes at low velocities would produce fractals, structure that appear invariant under changes of spatial scale (\cite {witten86}, \cite{meakin88} and \cite{donn1991}).  Since then, fractal aggregates have  been postulated to be the building blocks in protoplanetary disks (e.g., \cite{mukai1992}).  The formation of highly porous, fractal aggregates by dust grains colliding at low relative velocities ($< 0.2$ m/s was confirmed by experiments  (\cite{blumwurm2000}).  Finally, there is indirect evidence for porous aggregates in well-studied circumstellar disks like HR 4796A \cite{augereau1999}), and AU Microscopii (\cite{fitzgerald2007}).   

A defining characteristic of a fractal aggregate is that the average of the number of unit particles, $N$, within a radius $r$, is given by $N(r) = Ar^D$, where  $A$ is a constant, and $D$ is the fractal dimension.  From this relation it can be shown that the density  $\rho$ decreases with size according to the relation $\rho(r) \propto r^{D-3}$.  In the early  growth stage,  aggregates simply stick to each other,  and their density decreases accordingly (\cite{suyama08}).  These early aggregates have mass fractal dimension $D=1.75$ (\cite{katyal14}).  As the fractal aggregates grow, depending on the impact energy and the particle binding energy, the more filamentary structures are expected to break off and the aggregates become compressed, reaching a fractal dimension of $\sim 2.5 $ (\cite{suyama08}).  Even with collisional compression, these aggregates still retain very low densities of order $\lesssim 0.1$ kg m$^{-3}$, thanks to their fractal structure.  For icy aggregates growing beyond the snowline of protoplanetary disks, Okuzumi et al. (2012) found an even lower density, $\sim 10^{-2}$ kg m$^{-3}$, for a wide range of aggregate sizes, at the end of the collisional compression stage.  This result was echoed by Kataoka et al. (2013)  who found that sub-micron ice grains could form 100m-size planetesimals with density $10^{-2}$ kg m$^{-3}$ beyond 5 AU.   We note that fractal grains have been found in the coma of comet Churyumov-Gerasimenko in the form of fluffy aggregates (\cite{bentley16}), with densities $\lesssim  0.1$ kg m$^{-3}$ and tensile strength $T = 0.7$ kPa (\cite{fulle2015}).  

In this work we build on Moro-Martin's premise of a fractal structure for 'Oumuamua and interpret the observations in the fractal context. The paper is organized as follows.  In Section 2, we report a change in the rotation period of 'Oumuamua based on the available ground-based observations.  The change in rotation velocity is consistent with a very rapid response to the YORP effect, made possible only by the very small density of a fractal-like body.   We investigate the mechanical forces acting on 'Oumuamua due to centrifugal forces in Section 3, and the forces due to tidal forces in Section 4.

\section{YORP effect}


\begin{deluxetable*}{lcc}
\tablenum{1}
\tablecaption{Rotation Period of 'Oumuamua
\label{rotation}}
\tablewidth{0pt}
\tablehead{ \colhead{Work} & \colhead{Epoch of Last Exposure$^a$} & \colhead{$P$ [hr]} }

\startdata
Meech et al. (2017 & 53.25 & 7.34 \\
Drahus et al. (2017) & 54.52 & 7.55 \\
Bannister et al. (2017), Bolin et al. (2018) & 55.88 & 8.10 \\
Jewitt et al. (2017), in conjunction with Knight et al. (2017) & 56.29 & 8.26 \\
\enddata

\tablenotetext{a}{Epoch of last exposure of observing run, in MJD  = JD - 2458000.5}
\end{deluxetable*}

\begin{figure}[h!]
\plotone{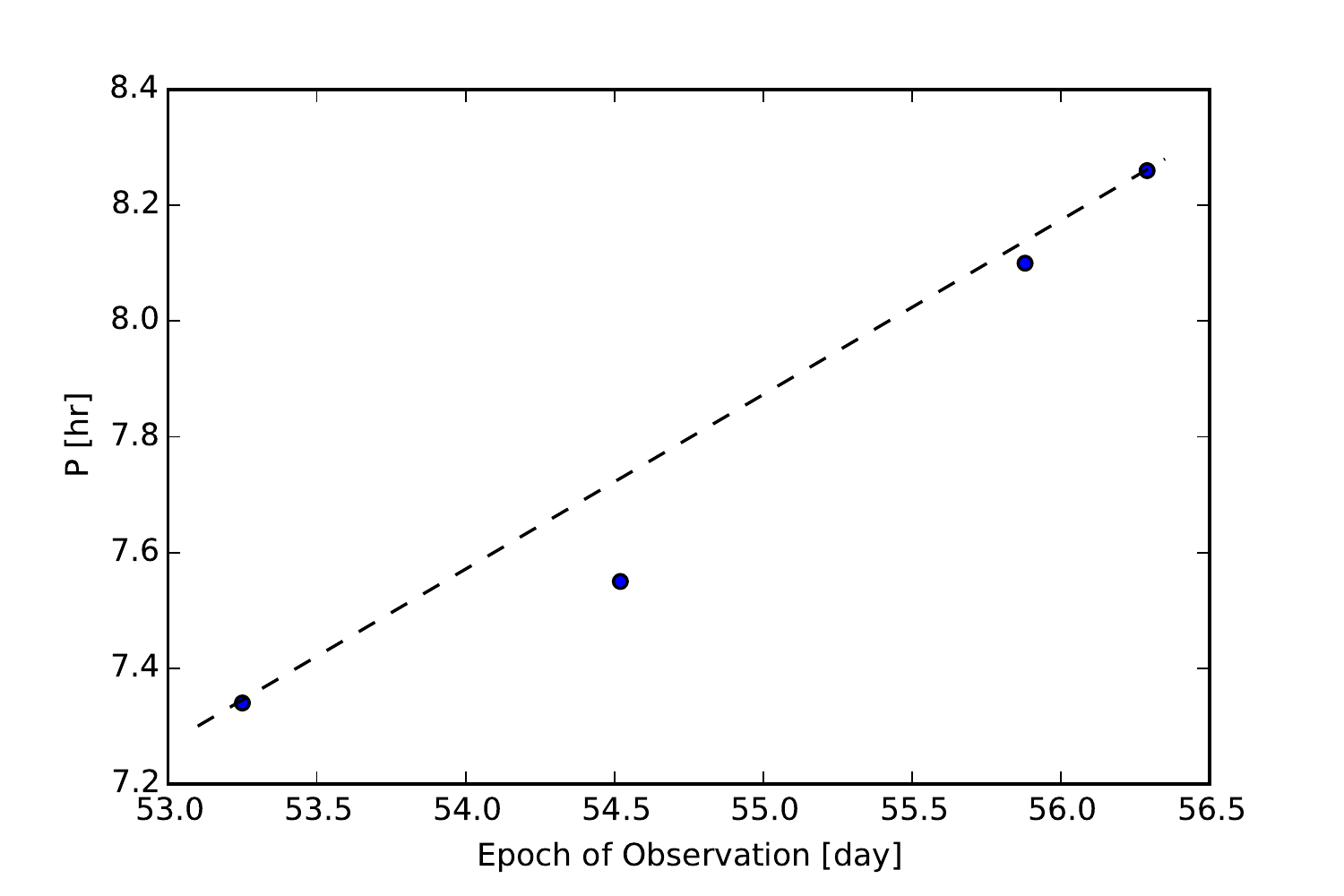}
\caption{'Oumuamua's reported rotation period as a function of epoch of observations  (in MJD = JD - 2458000.5).  The data are as listed in Table 1 and described in the text.  The plotted line simply connects the first and last data points, to guide the eye. \label{dPdt}}. 
\end{figure}

A summary of all the available optical photometry of 'Oumuamua can be found in \cite{belton18}  (their Table 1).  The ground-based data come from six different works, listed here in chronological order:  Meech et al. (2017), Drahus et al. (2017),  Bolin et al. (2018),  Bannister et al. (2017),  Knight et al. (2017), and Jewitt et al. (2017).    The Bannister/Bolin data can be combined as one set of data since they are very close in time (Belton et al.'s Figure 1) and both report the same rotation period.  Also, the Jewitt and Knight data can also be combined into one one set of data, since they overlap, and both sets of data were combined to obtain a rotation period (\cite{jewitt2017}).   With these combinations, there are 4 separate epochs of observations, reporting 4 different rotation periods.  These  rotation periods are listed in (our) Table 1, along with the epoch of the last exposure of each run, as reported by Belton et al.'s Table 1.  The data are also plotted in Figure (\ref{dPdt}).  The data indicate that the rotation period $P$ increases with time, with a rate of $dP/dt \sim \Delta P / \Delta t = 0.3$ hour/day, or $d\omega/dt =  10^{-10}$ s$^{-2}$.   The timescale required to change the rotation period is $\tau_s = P/(dP/dt) = 7.34 / 0.3 = 24$ days, where we used the earliest measured rotation period $P = 7.34$ hr (see Table 1).

If 'Oumuamua was propelled by radiation pressure, it is a natural question to ask whether radiation pressure also affected its rotational properties.  Small solar system bodies with low albedos are susceptible to the YORP effect, which refers to the torque that results when thermal radiation is anisotropically emitted from the object's surface (\cite{rubincam2000}).  The YORP characteristic timescale is a complex function of the object's shape, thermal properties, and spin axis, none of which is known for 'Oumuamua, so we estimate this timescale by scaling from previous observations of asteroids \cite{jewitt2015}, acknowledging the significant uncertainty.  Adapting Equation (3) of Jewitt et al., the YORP timescale is given by
\begin{eqnarray}
\tau_{YORP} [\mathrm{Myr}] \sim 4 \ \lbrack \frac{{r}}{1 ~ \mathrm{km}} \rbrack ^2 \ \lbrack \frac{R}{3 ~ {\mathrm{AU}}} \rbrack ^2 \ \lbrack \frac{\omega}{3 \times 10^{-4}~\mathrm{s^{-1}}} \rbrack   \nonumber \\  
\times\lbrack \frac{\orho}{2 \times 10^3 ~ \mathrm{kg m^{-3}}} \rbrack, 
\label{t_YORP}
\end{eqnarray}

\noindent where $r$ is the object's radius, $R$ is the heliocentric distance, $\omega = {2 \pi}/{P}$ is the angular velocity, and $\orho$ is the average bulk density. 
Adopting the values $r = 230$ m (see section 3 and \cite{rafikov18}), $R = 1$ AU, $\omega = 2 \times 10^{-4}$ s$^{-1}$, and the average density $\orho = 10^{-2}$ kg m$^{-3}$ (\cite{okuzumi12}, \cite{kataoka2013}, we obtain $\tau_{YORP} \sim 29$ days.   We note that this is completely consistent with the timescale required to change the rotation period, $\tau_s$.  This close agreement must be somewhat of a coincidence, considering the uncertainties involved in Equation (1), but it does show that the observed $dP/dt$ is consistent with the YORP effect, provided the bulk density is very low.  It thus takes only $\sim 1$ month for the YORP effect to significantly change the rotation speed of 'Oumuamua.  This is very short compared to the $\sim$ 1 Myr timescale for asteroids, and is made possible only by 'Oumuamua's extraordinarily small density of $\orho = 10^{-2}$ kg m$^{-3}$, compared to the density $\sim 10^3$ kg m$^{-3}$ for most asteroids.   We conclude that 'Oumuamua's observed spin-down is plausibly attributed to the YORP torque acting on a fractal body with a very small bulk density, corresponding to a large area-to-mass ratio.  

\section{Mechanics of a rotating fractal}

There are considerable uncertainties in most of the physical parameters measured for 'Oumuamua.  Hereafter we adopt the following parameters: the object has the dimensions $a \times b  \times c = $ 230 m x 35 m x 35 m (\cite{rafikov18}), mass $(4/3) \pi \orho  abc = 1.2 \times 10^4$ kg, and rotates with a rotation period of $P=$ 8.26 hour.  The large brightness variations in 'Oumuamua's lightcurve most likely arise from rotation around the short axis, i.e., the axis of maximum inertia.    Given 'Oumuamua's susceptibility to radiation pressure torques, this spin axis may precess, as might the long axis (\cite{belton18}), but in this paper we shall only assume rotation about the short axis.  

Leaving tidal forces and the forces exerted by the radiation aside for now we address the question of the forces created by the rotation around the center of mass itself.  

\subsection{Fractal properties and particle size}
A fractal is a structure that is scale invariant. In case of 'Oumuamua the structure is assumed to have formed through random aggregation processes and will thus be disordered. The scale invariance implies that it will have holes of all length scales inside. However, the object would break apart if a hole would be
bigger in linear size than $b$, and we will thus take $b$ to be the upper limit of the fractal behavior (a useful conceptual comparison would be that of a train of Sierpinsky gaskets).
\begin{figure}[h!]
 \plotone{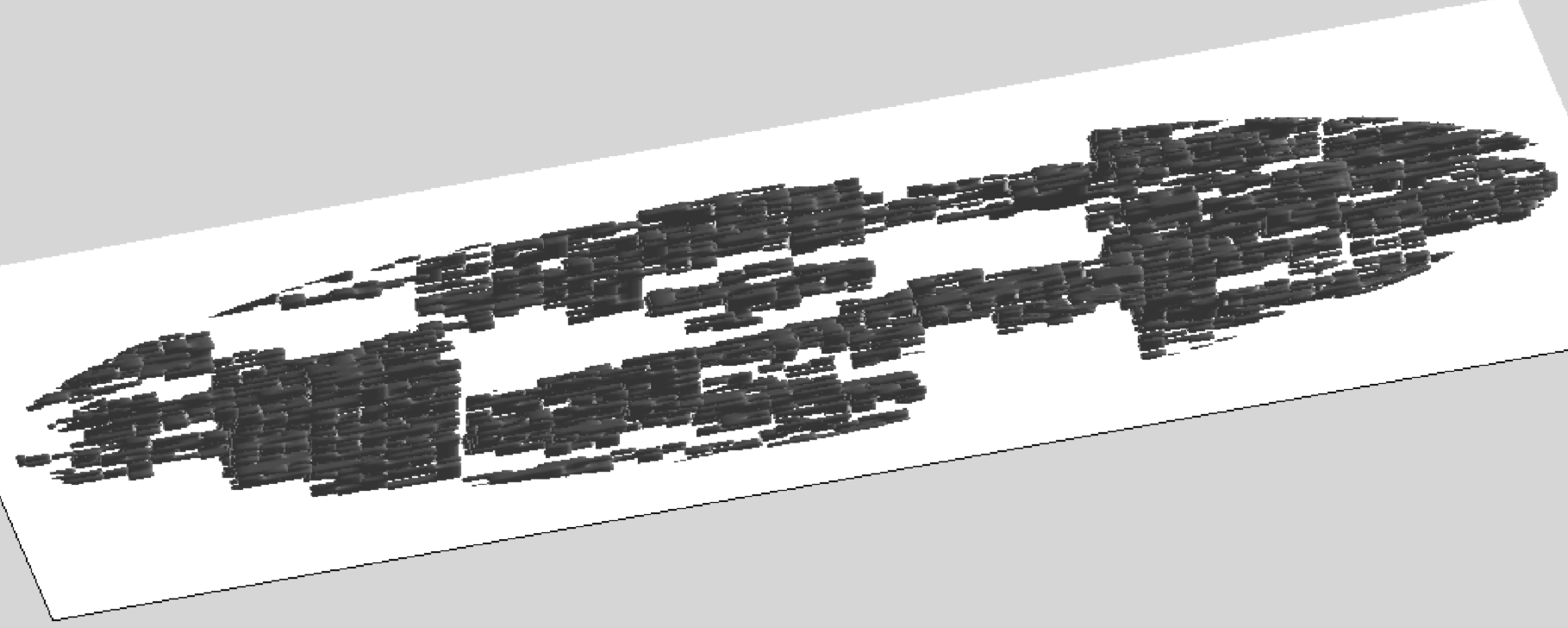}
\caption{ A fractal of the same dimension $D$ and aspect ratio as is assumed for  'Oumuamua.} \label{concept} 
\end{figure}
In Figure \ref{concept} a fractal of dimension $D=$2.35 is shown. It  is constructed by partitioning the body into cubes, and each cube is then partitioned into 3$^3$ sub-cubes. 
We then remove  sub-cubes from every cube  at random so as to keep a given average fraction $f=3^D/3^3$  
of cubes. The remaining cubes are treated the same way until the a  fractal of dimension $D$  results. 
This simple algorithm does {\it not} correspond to the physical aggregation processes that may have formed 'Oumuamua, 
but it serves as a conceptual guide and illustrates the degree of surface fluctuations and transparency that is typical of such a disordered fractal. 

The mass of this fractal may then be written
\be
m(r) = m_0  \left( \frac{r}{r_0}\right)^D
\ee
where $r$ is the distance from a point on the fractal, $r_0$  and $m_0$ the smallest, or elementary,  particle radius and mass, and $D$ the fractal
dimension. 

The density of such a fractal is
\be  
\rho (r) = \frac{3}{4\pi r^3} m(r) = \rho_0  \left(
  \frac{r}{r_0}\right)^{D-3}
\label{density}
\ee  
where  $\rho_0$ is the density of the material making up the smallest particles. We shall use the density of ice, $\rho_0 =10^3$ kg$/$m$^3$.

The particle size follows by identifying the observed  $\orho=$ 10$^{-2}$ kg $/$m$^3$ with $\rho (b)$.
Solving \eq{density} for $D$ gives 
\be
D = 3 + \frac{\ln (\rho (b) / \rho_0)}{\ln (b/r_0)}
\label{D} .
\ee

\begin{figure}[h!]
\plotone{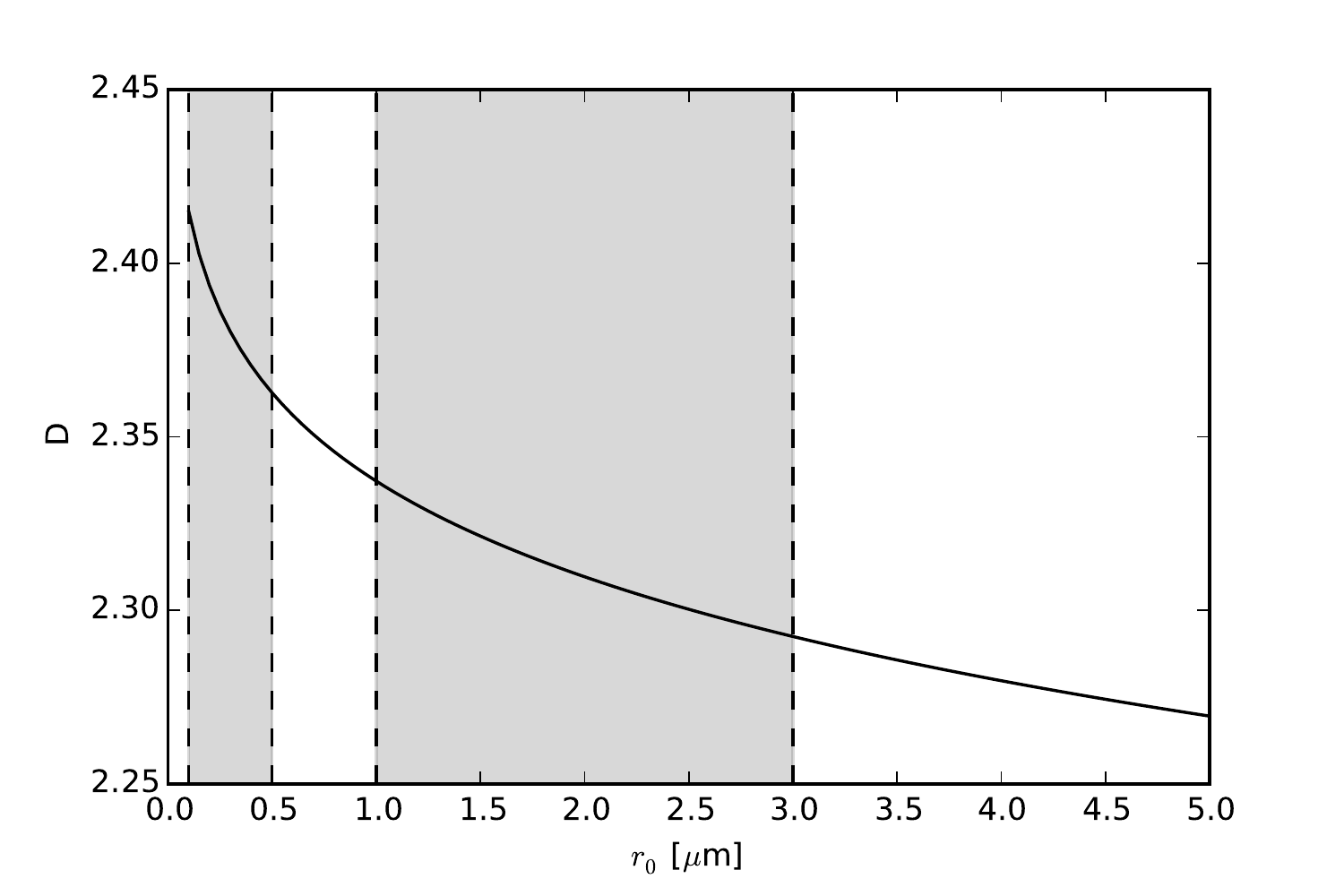}
\caption{'Fractal dimension $D$ as a function of $r_0$.  The shaded areas denote the range of $r_0$ measured in comets Wild 2 and Churyumov-Gerasimenko.  \label{D_vs_r0}}.
\end{figure}

\noindent Measurements of cometary fractal aggregates are available for comets Wild 2 (\cite{kearsley2008}) and Churyumov-Gerasimenko (\cite{bentley16}).  The smallest grains of Comet Wild 2 have size $\lesssim 500$ nm (\cite{kearsley2008}), while those of comet Churyumov-Gerasimenko have sizes in the range $260 - 500$ nm, and $1-3 \micron$ (\cite{bentley16}).  Using the representative sizes as $r_0 = 100$ nm and $1\mu$m, we can calculate $D$ using \eq{D}.  Figure (\ref{D_vs_r0}) shows $D$ as a function of $r_0$.  Within the range of measured $r_0$, \eq{D}  predicts that 
$D$ lies in the range $2.3 \leq D \leq 2.4$ in agreement with Figure 1 in \cite {moromartin19b}. 
The $D$-value is comparable to, but somewhat smaller than the  $D \sim 2.5$ predicted by numerical simulations of aggregates that have undergone collisional deformation (\cite{suyama08}, \cite{okuzumi12}).  Note that due to the factor $(b/r_0)$,   the fractal behavior extends over a full 8 orders of magnitude.  This extraordinary range of length scales over which the structure appears to be fractal is much larger than the range of scales normally observed on Earth (typically $\le 3$, \cite{feder88}), or in other extraterrestrial aggregate structures like cometary grains (\cite{bentley16}) and interplanetary dust particles (\cite{rietmeijer1998}).

In the following we shall use $r_0=$ 1 $\mu$m whenever a concrete number is needed, but as we will see, several key values do not depend on $r_0$.

\subsection{Mechanical forces due to rotation}

In this section we consider 4 different stresses:  (1) $\osigma_c$, the averaged stress resulting from  centrifugal forces, (2) $\sigma_{link}$, the stress from centrifugal forces over a single link, (3) $\sigma_{yield}$, the yield stress for a single link   and (4) $\osigma_{yield}$, the  yield stress averaged over the whole body.

Since we are considering the short axis rotation of an elongated object, the principal axis of the strongest tensile stress will
largely point along the long axis, i.e., the forces will act across planes of area $\pi b^2$, the smallest plane that  intersects the structure. 
First, we consider the average stress $\osigma_c$  over such a plane. We will take a coordinate $z$ to point along the long axis and taking its
origin in the centre of mass. Averaging across the plane normal to $z$, the balance between the centrifugal force and this stress may then be written

\be
\frac{d \osigma_c}{dz} = -\orho \omega^2 z
\ee
where $\omega = 2\pi /P =$ 2 10$^{-4}$ Hz. Using the boundary condition
$\osigma_c (a)=0$ the above equation may be integrated to yield the
internal stress
\be
\osigma_c (z) = \frac{\orho}{2} \omega^2 ( a^2 - z^2)
\ee
which is maximum for $z=0$, or
\be
\osigma_{c\; {max}}
= \frac{\orho}{2} \omega^2 a^2 = 10^{-5} \mbox{Pa}, 
\label{ugytd}
\ee

\noindent using the values $\bar{\rho} = 10^{-2}$ kg m$^{-3}, a = 230$ m, and $\omega = 2 \times 10^{-4}$ s$^{-1}$.  We note that this is within a factor of 2 of \cite{bialy18}'s estimate of the stress due to centrifugal forces.  We are aware that inside a fractal, some bonds may not share in the stresses. In fact, the backbone (the load-bearing structure) may indeed have a smaller fractal dimension than the rest of the structure; this will be discussed later, in Section 3.4.

\subsection{Estimate of internal mechanical strength}

Assuming for now that all bonds share equally in the load, we can calculate the stress $\sigma_{link}$ along links of particles, based on the fact that 'Oumuamua can survive  centrifugal forces.   We equate the following expression for the tensile force
\be
\pi b^2 \osigma_c = \phi  \pi b^2 \sigma_{link}
\ee
where $\phi = \orho /\rho_0 = 10^{-5}$ is the solid fraction.  This gives the  link stress
\be
\sigma_{link} = \frac{\rho_0}{\orho} \osigma_c
= \frac{\rho_0}{2} \omega^2 a^2 = 1 \mbox{Pa} ,
\ee
which is rather modest. Note that the expression for $\sigma_{link}$ does not depend on $r_0$ and $D$.

We can also calculate the upper limit of the link stress before breaking.  This estimate is based on the physical picture that the particles have a surface energy per unit area $\gamma$ that is converted to elastic energy as  the particles connect and deform (\cite{johnson71}). The force $F_S$ needed to separate the particles again is estimated as the surface energy divided by the distance characterizing the deformation. Their result
\be
F_S=\frac{\pi \gamma r_0}{2}
\ee

\noindent is notably independent of such material parameters as Young's modulus and Poisson ratio,  and the surface energy could be due to any kind of interatomic forces.  It leads to the following prediction of the critical stress

\be
\sigma_{yield} = \frac{F_S}{\pi r_0^2} = \frac{\gamma}{2r_0} ,
\label{sigmac}
\ee
which is inversely proportional to $r_0$.  We note that \cite{suyama08} also used this same expression for the force where the surface energy is taken to be that of ice $\gamma = 0.1$ J$/$m$^2$.  This gives

\be
\sigma_{yield}= \frac{\mbox{2.5 $\mu$m}}{r_0} \mbox{20 kPa} .
\ee

With $r_0=$1$\mu$m this critical stress is almost 4 orders of magnitude larger than $\sigma_{link}$, so the constant load sharing assumption
certainly predicts a stable structure.

We may also inquire about the yield stress $\osigma_{yield}$ that will cause the whole structure to break apart. This will happen when
\be
 \osigma_{yield}
 = \frac{\orho}{\rho_0} \sigma_{yield}
 = \mbox{0.2 Pa} .
\ee 
This means that any external stress acting on the whole structure and exceeding this small value will deform or break it.
The single link tensile strength $\sigma_{yield}$ is much larger than the actual link stress $\sigma_{link}$ because of the relatively slow rotation rate.
Yet, due to the very low density, the corresponding average $\osigma_{yield}$ is quite small. In fact,  $\osigma_{yield}$ is more than an order of magnitude
smaller than the stress that would result from Earth's gravity, $\osigma_{gravity}=  mg/(\pi a b) = $3 Pa.  In other words, 'Oumuamua would simply collapse on Earth.  It should be noted that $ \osigma_{yield}$ is still much larger than the radiation pressure which at a distance of 0.25 AU is about 80 $\mu$Pa.

\subsection{What if the backbone dimension is smaller than D?}
So far we have assumed that the entire fractal structure carries an equal load. It is well known, however, that most random fractals have dangling ends that do not carry any load.  In granular packings, external compression gives rise to ramified force networks through the grain assembly.  These force networks strongly dependent on the nature and orientation of the external loading.  It is then to be expected  that 'Oumuamua, which must already have experienced some deformation during its aggregation process, might also deform according to mechanical forces caused by its rotational motion. 
The result will be some kind of backbone structure that carries most of the load.

Since the initial structure is assumed to be a fractal, with dangling substructures on all scales smaller than the fractal itself, it is reasonable to characterize the backbone as a fractal too, with  a fractal dimension $D_B$ that is smaller than that of the  original structure (\cite{stauffer85}).

As before, we also assume equal load sharing, but now inside the backbone only. In this way we may inquire whether the presence of dangling ends on all scales would catastrophically disrupt the structure.
 The larger the fraction of dangling substructures that does not  take part in the load carrying, the larger the stress on the remaining backbone structure.  Thus we can calculate a lower bound on the backbone fractal dimension $D_B$, below which
the bond stress becomes larger than $\sigma_{yield}$.

The backbone will have a mass density $\rho_B$ defined in exactly the same way as that leading to \eq{density}, except with a different fractal dimension:
\be 
\rho_B (r) =  \rho_0  \left(
  \frac{r}{r_0}\right)^{D_B-3} 
\label{densityB}
\ee  

\noindent The volume fraction of the backbone averaged over a cross-section of area $\pi b^2$ is then

\be 
\phi_B = \frac{\rho_B (b)}{\rho_0}
=   \left(  \frac{b}{r_0}\right)^{D_B-3} .
\label{fracB}
\ee 

\noindent When the forces are distributed only on the backbone, we must make the replacement
\be
\phi \rightarrow  \phi_B
\ee

so that now the link stress becomes

\be
\sigma_{link\; B} = \frac{ \osigma_c}{\phi_B} =
\frac{\rho_0}{\orho }
\frac{\orho}{\rho_B(b) } \osigma_c =
\sigma_{link}
\left(   \frac{b}{r_0}\right)^{D-D_B} .
\ee

Requiring that $\sigma_{link\; B} < \sigma_{yield}$ we get the condition
\be
D_{B}
>  D - \frac{\ln \left( \frac{ \sigma_{yield}    }{\sigma_{link}}\right) }{\ln \left( \frac{b}{r_0}\right)},
\ee
which, when $r_0 = 1 \mu$m, gives the condition $D _B>  D - 0.62$.

From a purely geometrical point of view this is rather a conservative condition. The mass ratio between the  load carrying minimal backbone and the total structure is then $(r_0/b)^{D-D_{B{min}}} =(r_0/b)^{0.62} = 2 \times 10^{-5}$,
which is indeed a small fraction.
Since stretching of a structure with only a few load carrying parts will tend to redistribute the force over other parts, it is
reasonable to assume that the mass fraction of load carrying bonds will increase well before the entire system breaks.

\section{Tidal forces}

Considering that 'Oumuamua came within 0.25 AU of the Sun, we now consider the effect of  tidal forces.
\subsection{The internal stresses caused by tidal forces}
Since we have already considered the internal forces due to the rotation around its center of mass, we now consider the motion of an elongated object 
moving around the sun with its long axis constantly pointing towards the sun. 
We then estimate the maximum stress due to tidal forces by the difference in the gravitational forces $F(r)$  on the closest and most distant point of our object, that is
\be 
\osigma_{t\; {max}} = \frac{\Delta F(r)
}{\pi b^2}.
\label{jufuf}
\ee
Here $F(r) = \Gamma Mm/r^2$ where $\Gamma$ is the gravitational constant, $M$ the sun mass and $m$ the total mass of 'Oumuamua.
Taking $R$ to be the center of mass distance from the sun and
$\Delta F(r)$ to be calculated between the points $r=R \pm a$ this   means that $\Delta F(r) = - 4(a/R) F(R)$.  For the sake of getting an estimate,
we may approximate the 'Oumuamua motion  by a circular one with an orbit angular velocity $\Omega$. Then the balance between
the centripetal force  and gravity may be written $F(R) = m \Omega^2 R$, giving  $\Delta F = -4 m \Omega^2 a$.  
Inserting this in \eq{jufuf}  and using $\orho =m/(\pi ab^2)$ we can write
\be 
\osigma_{t\; {max}}
= 4 \orho \Omega^2 a^2 . 
\ee
Comparing this with the stress due to the rotation around the center of mass given in \eq{ugytd} we can write
\be
\frac{\osigma_{t\; {max}}
}{\osigma_{c \; {max}}} = 4 \frac{\Omega^2}{\omega^2 } \ll 1 . 
\ee
In other words, tidal stresses are smaller than the rotational stress by a factor that is the square of $P$ divided by the
 time for a hypothetical circular orbit around the sun. Since we have already shown that $\osigma_{c \; {max}}$ is unlikely to destabilize 'Oumuamua,
the same may clearly be said for the role of   $\osigma_{t\; {max}}$.

\subsection{The torque due to tidal forces}
Taking 'Oumuamua now
to a have an arbitrary shape
the torque around the center of mass may be written
\be 
\btau = - \int dV \rho (\br )  \bg \times \br 
\ee
where $\br $ is measured from the center of mass, so that $\int dV
\rho  \br
=0$.  The acceleration of gravity is 
\be
\bg (\br ) =- \frac{\Gamma M}{|   \bR + \br |^3} ( \bR + \br ) .
\ee
Here $\bR$ is  the center of mass position relative to the sun, we may be split $\bg$  into a constant and variable part $\bg (\br ) = \bg_0 + \Delta \bg$, where $\bg_0 =
\bg (\bR )$, so that the torque 
\be 
\btau = - \bg_0 \times \int dV \rho (\br )   \br  +  \int dV \rho (\br  
)  \Delta \bg  \times \br .  
\ee  
Here, the first term vanishes and $\Delta \bg = \br \cdot \nabla \bg
(\br )$ may be written
\be
\Delta \bg = \frac{\Gamma M}{|\bR + \br |^3
} \left[
  \br - 3 \frac{
\br \cdot ( \bR + \br ) ( \bR + \br )
  }{|\bR + \br |^2}\right] .
\ee
Keeping only  terms to leading order in $r/R$ 
\be
\Delta \bg = \frac{\Gamma M}{R^3} \left\{
    \br - 3 {
      \bu \cdot \br \bu  }\left( 1 + o \left( \frac{r}{R} \right) \right)
  \right\}
\ee
where we have introduced the constant unit vector $\bu = \bR /R$.
Now, since $\br \times \br =0$
\be 
\btau 
= \frac{3\Gamma M}{R^3} \int dv \rho
( \bu_R \cdot \br ) \bu_R \times \br \left( 1 + o \left( \frac{r}{R}
  \right) \right) .
\ee
Introducing the moment of inertia tensor
\be
T_{ij} = \int dV \rho r_i r_j,
\ee
we may write  the above equation on component form as
\be
\tau_i = \epsilon_{ijk} \frac{3 \Gamma M}{R^3} T_{nk} u_j u_n 
+ o\left(
\int dV \rho \Gamma M \frac{r^3}{R^4}
\right) 
\label{kjhgyt}
\ee
where we have introduced the antisymmetric Levi-Civita tensor and neglected the vectorial nature of the last term since we are only interested in its order  of magnitude. 

Since $T_{nk}$ is symmetric we may assume that we are already in the orthogonal coordinate system where it is also diagonal so that $T_{nk}
\propto \delta_{nk} $, where $\delta_{nk}$  is the Kronecker delta. 
This implies that
\be
\epsilon_{ijk}  T_{nk} u_j u_n \propto  \epsilon_{ijk}  u_j u_k =0
\ee
since the $\epsilon$-term is antisymmetric in $jk$ and the last terms
are
symmetric in $jk$. Then we are only left with the last contribution 
to the integral of \eq{kjhgyt},  and
\be
\tau \approx   \int dV \rho \Gamma M \frac{r^3}{R^4} \sim
m g_0 \frac{ a^3}{R^2}
\label{kjhhgyt}
\ee
where $m$ is the mass of Oumuamua.

We may now estimate the characteristic time of change of angular momentum
\be 
t_{tidal} = \frac{I\omega }{\tau}
\ee
where $I\sim m a^2$  is the moment of inertia and $\omega =2$ $10^{-4}$s$^{-1}$ the angular velocity of Oumuamua.  
Using this and 
\eq{kjhhgyt} gives
\be 
t_{tidal} = \frac{I\omega }{\tau} = \frac{\omega}{g_0} \frac{R^2}{a}
\approx \mbox{10$^9$ years} .
\ee
So, the effect of tidal forces on the angular velocity of Oumuamua  during its passage through the inner solar system should be quite negligible. 

\section{Conclusion}

We report observations and calculations that support the hypothesis of  a fractal structure of 'Oumuamua:

(1) Ground-based observations show that the rotation period of 'Oumuamua increased at the rate of $dP/dt \sim 0.3$ hour/day, corresponding to a timescale of $\sim 24$ days to change the angular momentum.  This timescale is consistent with the YORP characteristic timescale, provided the bulk density is $\orho \sim 10^{-2}$ kg/m$^{-3}$, as expected from a fractal structure (\cite{okuzumi12}, \cite{katoaka13},  \cite{moromartin19b}).  

(2) The expected yield stresses inside a fractal structure like 'Oumuamua are far larger than the actual mechanical stresses
caused by the orbital and rotational  motion, even though the structure could not survive the gravity on Earth's surface. Thus, even though our fractal structure
is very fragile, it should not break apart in its trajectory through the inner Solar System.  If the internal load is evenly distributed,  the stress due to centrifugal forces is 4-5 orders of magnitude smaller than the cohesion force between particles that make up the body (\cite{johnson71}).
However, this result holds even if the mechanical loads are carried by a backbone with a lower fractal dimension and there are dangling substructures with no load.

(3) Assuming that  the smallest grain size is comparable to that found in comets (e. g., comet 67P/Churyumov-Gerasimenko and comet Wild 2), the average bulk density implies  a fractal dimension $2.3 \leq D \leq 2.4$.  'Oumuamua thus appears to display a fractal structure over 8 orders of magnitude, a much larger range of length scale than that previously seen on Earth or in space.

\acknowledgments
We thank the Research Council of Norway through its Centres of Excellence funding scheme, project number 262644, and the LIA France-Norway D-FFRACT.

\bibliography{all}

\begin{thebibliography}{}
\expandafter\ifx\csname natexlab\endcsname\relax\def\natexlab#1{#1}\fi
\providecommand{\url}[1]{\href{#1}{#1}}
\providecommand{\dodoi}[1]{doi:~\href{http://doi.org/#1}{\nolinkurl{#1}}}
\providecommand{\doeprint}[1]{\href{http://ascl.net/#1}{\nolinkurl{http://ascl.net/#1}}}
\providecommand{\doarXiv}[1]{\href{https://arxiv.org/abs/#1}{\nolinkurl{https://arxiv.org/abs/#1}}}

\bibitem[{{Augereau} {et~al.}(1999){Augereau}, {Lagrange}, {Mouillet},
  {Papaloizou}, \& {Grorod}}]{augereau1999}
{Augereau}, J.~C., {Lagrange}, A.~M., {Mouillet}, D., {Papaloizou}, J.~C.~B.,
  \& {Grorod}, P.~A. 1999, \aap, 348, 557

\bibitem[{Bannister {et~al.}(2017)Bannister, Schwamb, Fraser, Marsset,
  Fitzsimmons, Benecchi, Lacerda, Pike, Kavelaars, Smith, Stewart, Wang, \&
  Lehner}]{bannister2017}
Bannister, M.~T., Schwamb, M.~E., Fraser, W.~C., {et~al.} 2017, The
  Astrophysical Journal, 851, L38

\bibitem[{Belton {et~al.}(2018)Belton, Hainaut, Meech, \& Mueller}]{belton18}
Belton, M. J.~S., Hainaut, O.~R., Meech, K.~J., \& Mueller, B. 2018, Astrophys.
  J. Lett., 856

\bibitem[{Bentley {et~al.}(2016)Bentley, Schmied, Mannel, Torkar, Jeszenszky,
  Romstedt, Levasseur-Regourd, Weber, Jessberger, Ehrenfreund, Koeberl, \&
  Havnes}]{bentley16}
Bentley, M.~S., Schmied, R., Mannel, T., {et~al.} 2016, Nature, 537, 73

\bibitem[{Bialy \& Loeb(2018)}]{bialy18}
Bialy, S., \& Loeb, A. 2018, Astrophys. J. Lett., 868, 5

\bibitem[{{Blum} \& {Wurm}(2000)}]{blumwurm2000}
{Blum}, J., \& {Wurm}, G. 2000, \icarus, 143, 138

\bibitem[{Bolin {et~al.}(2017)Bolin, Weaver, Fernandez, Lisse, Huppenkothen,
  Jones, Juri{\'{c}}, Moeyens, Schambeau, Slater, Ivezi{\'{c}}, \&
  Connolly}]{bolin2017}
Bolin, B.~T., Weaver, H.~A., Fernandez, Y.~R., {et~al.} 2017, The Astrophysical
  Journal, 852, L2

\bibitem[{{Do} {et~al.}(2018){Do}, {Tucker}, \& {Tonry}}]{do2018}
{Do}, A., {Tucker}, M.~A., \& {Tonry}, J. 2018, \apjl, 855, L10

\bibitem[{{Donn}(1991)}]{donn1991}
{Donn}, B. 1991, in Astrophysics and Space Science Library, Vol. 167, IAU
  Colloq. 116: Comets in the post-Halley era, ed. J.~{Newburn}, R.~L.,
  M.~{Neugebauer}, \& J.~{Rahe}, 335

\bibitem[{Drahus {et~al.}(2018)Drahus, Guzik, Waniak, Handzlik, Kurowski, \&
  Xu}]{drahus2018}
Drahus, M., Guzik, P., Waniak, W., {et~al.} 2018, Nature Astronomy, 2, 407

\bibitem[{Feder(1988)}]{feder88}
Feder, J. 1988, Fractals (New York: Plenum Press)

\bibitem[{{Fitzgerald} {et~al.}(2007){Fitzgerald}, {Kalas}, {Duch{\^e}ne},
  {Pinte}, \& {Graham}}]{fitzgerald2007}
{Fitzgerald}, M.~P., {Kalas}, P.~G., {Duch{\^e}ne}, G., {Pinte}, C., \&
  {Graham}, J.~R. 2007, \apj, 670, 536

\bibitem[{Fraser {et~al.}(2018)Fraser, Pravec, Fitzsimmons, Lacerda, Bannister,
  Snodgrass, \& Smoli{\'c}}]{fraser2018}
Fraser, W.~C., Pravec, P., Fitzsimmons, A., {et~al.} 2018, Nature Astronomy, 2,
  383

\bibitem[{{Fulle} {et~al.}(2015){Fulle}, {Della Corte}, {Rotundi}, {Weissman},
  {Juhasz}, {Szego}, {Sordini}, {Ferrari}, {Ivanovski}, {Lucarelli}, {Accolla},
  {Merouane}, {Zakharov}, {Mazzotta Epifani}, {L{\'o}pez-Moreno},
  {Rodr{\'\i}guez}, {Colangeli}, {Palumbo}, {Gr{\"u}n}, {Hilchenbach},
  {Bussoletti}, {Esposito}, {Green}, {Lamy}, {McDonnell}, {Mennella}, {Molina},
  {Morales}, {Moreno}, {Ortiz}, {Palomba}, {Rodrigo}, {Zarnecki}, {Cosi},
  {Giovane}, {Gustafson}, {Herranz}, {Jer{\'o}nimo}, {Leese},
  {L{\'o}pez-Jim{\'e}nez}, \& {Altobelli}}]{fulle2015}
{Fulle}, M., {Della Corte}, V., {Rotundi}, A., {et~al.} 2015, \apjl, 802, L12

\bibitem[{{Gaidos} {et~al.}(2017){Gaidos}, {Williams}, \& {Kraus}}]{gaidos17}
{Gaidos}, E., {Williams}, J., \& {Kraus}, A. 2017, Research Notes of the
  American Astronomical Society, 1, 13

\bibitem[{{Jewitt} {et~al.}(2015){Jewitt}, {Hsieh}, \& {Agarwal}}]{jewitt2015}
{Jewitt}, D., {Hsieh}, H., \& {Agarwal}, J. 2015, The Active Asteroids
  (University of Arizona Press, Tucson), 221--241

\bibitem[{Jewitt {et~al.}(2017)Jewitt, Luu, Rajagopal, Kotulla, Ridgway, Liu,
  \& Augusteijn}]{jewitt2017}
Jewitt, D., Luu, J., Rajagopal, J., {et~al.} 2017, The Astrophysical Journal,
  850, L36

\bibitem[{Johnson {et~al.}(1971)Johnson, Kendall, \& Roberts}]{johnson71}
Johnson, K.~L., Kendall, K., \& Roberts, A.~D. 1971, Proc. Roy. Soc. London,
  324, 301

\bibitem[{{Kataoka} {et~al.}(2013){Kataoka}, {Tanaka}, {Okuzumi}, \&
  {Wada}}]{kataoka2013}
{Kataoka}, A., {Tanaka}, H., {Okuzumi}, S., \& {Wada}, K. 2013, \aap, 557, L4

\bibitem[{Kataoka {et~al.}(2013)Kataoka, Tanaka, Okuzumi, \& Wada}]{katoaka13}
Kataoka, A., Tanaka, H., Okuzumi, S., \& Wada, K. 2013, Astr. and Astrophys.,
  A4, 554

\bibitem[{Katyal {et~al.}(2014)Katyal, Banerjee, \& Sanjay}]{katyal14}
Katyal, N., Banerjee, V., \& Sanjay, P. 2014, J. Quant. Spectr. Rad. Transf.,
  146, 290

\bibitem[{{Kearsley} {et~al.}(2008){Kearsley}, {Borg}, {Graham}, {Burchell},
  {Cole}, {Leroux}, {Bridges}, {H{\"o}rz}, {Wozniakiewicz}, {Bland },
  {Bradley}, {Dai}, {Teslich}, {See}, {Hoppe}, {Heck}, {Huth}, {Stadermann},
  {Floss}, {Marhas}, {Stephan}, \& {Leitner}}]{kearsley2008}
{Kearsley}, A.~T., {Borg}, J., {Graham}, G.~A., {et~al.} 2008, Meteoritics and
  Planetary Science, 43, 41

\bibitem[{Knight {et~al.}(2017)Knight, Protopapa, Kelley, Farnham, Bauer,
  Bodewits, Feaga, \& Sunshine}]{knight2017}
Knight, M.~M., Protopapa, S., Kelley, M. S.~P., {et~al.} 2017, The
  Astrophysical Journal, 851, L31

\bibitem[{Meakin(1988)}]{meakin88}
Meakin, P. 1988, Ann. Rev. Mod. Chem., 39, 237

\bibitem[{Meech {et~al.}(2017)Meech, Weryk, Micheli, Kleyna, Hainaut, Jedicke,
  Wainscoat, Chambers, Keane, Petric, Denneau, Magnier, Berger, Huber,
  Flewelling, Waters, Schunova-Lilly, \& Chastel}]{meech2017}
Meech, K.~J., Weryk, R., Micheli, M., {et~al.} 2017, Nature, 552, 378 EP

\bibitem[{Micheli {et~al.}(2018)Micheli, Farnocchia, Meech, Buie, Hainaut,
  Prialnik, Schorghofer, Weaver, Chodas, Kleyna, Weryk, , Wainscoat, Ebeling,
  Keane, Chambers, D.Koschny, \& Petropoulos}]{micheli18}
Micheli, M., Farnocchia, D., Meech, K., {et~al.} 2018, Nature

\bibitem[{{Moro-Mart{\'\i}n}(2018)}]{moromartin18}
{Moro-Mart{\'\i}n}, A. 2018, \apj, 866, 131

\bibitem[{{Moro-Mart{\'\i}n}(2019{\natexlab{a}})}]{moromartin19a}
---. 2019{\natexlab{a}}, \aj, 157, 86

\bibitem[{{Moro-Mart{\'\i}n}(2019{\natexlab{b}})}]{moromartin19b}
---. 2019{\natexlab{b}}, \apjl, 872, L32

\bibitem[{{Mukai} {et~al.}(1992){Mukai}, {Ishimoto}, {Kozasa}, {Blum}, \&
  {Greenberg}}]{mukai1992}
{Mukai}, T., {Ishimoto}, H., {Kozasa}, T., {Blum}, J., \& {Greenberg}, J.~M.
  1992, \aap, 262, 315

\bibitem[{Okuzumi {et~al.}(2012)Okuzumi, Tanaka, Kobayashi, \&
  Wada}]{okuzumi12}
Okuzumi, S., Tanaka, H., Kobayashi, H., \& Wada, K. 2012, Astrophys. J., 106,
  752

\bibitem[{Rafikov(2018)}]{rafikov18}
Rafikov, R. 2018, Astrophys. J. Lett., 867, 17

\bibitem[{Rietmeijer \& Nuth(1998)}]{rietmeijer1998}
Rietmeijer, F.~J., \& Nuth, J.~A. 1998, Earth, Moon, and Planets, 82, 325

\bibitem[{{Rubincam}(2000)}]{rubincam2000}
{Rubincam}, D.~P. 2000, \icarus, 148, 2

\bibitem[{{Seligman} {et~al.}(2019){Seligman}, {Laughlin}, \&
  {Batygin}}]{seligman19}
{Seligman}, D., {Laughlin}, G., \& {Batygin}, K. 2019, \apjl, 876, L26

\bibitem[{Stauffer \& Aharony(1985)}]{stauffer85}
Stauffer, D., \& Aharony, A. 1985, Introduction to percolation theory, 3rd edn.
  (Taylor and Francis)

\bibitem[{Suyama {et~al.}(2008)Suyama, K.Wada, \& Tanaka}]{suyama08}
Suyama, T., K.Wada, \& Tanaka, H. 2008, Astrophys. J., 684, 1310

\bibitem[{{Trilling} {et~al.}(2018){Trilling}, {Mommert}, {Hora}, {Farnocchia},
  {Chodas}, {Giorgini}, {Smith}, {Carey}, {Lisse}, {Werner}, {McNeill},
  {Chesley}, {Emery}, {Fazio}, {Fernandez}, {Harris}, {Marengo}, {Mueller},
  {Roegge}, {Smith}, {Weaver}, {Meech}, \& {Micheli}}]{trilling18}
{Trilling}, D.~E., {Mommert}, M., {Hora}, J.~L., {et~al.} 2018, \aj, 156, 261

\bibitem[{Witten \& Cates(1986)}]{witten86}
Witten, T., \& Cates, M.~E. 1986, Science, 232, 4758

\end{thebibliography}
\bibliographystyle{aasjournal}
\end{document}